\documentclass[twocolumn,superscriptaddress,nofootinbib]{revtex4-1}
\usepackage{graphicx}
\usepackage{color}
\usepackage{amsmath}
\usepackage{lineno}
\usepackage{hyperref}
 \usepackage{relsize}
\usepackage[table]{xcolor}
\begin{document}
\title{
The balance of attractive and repulsive hadronic interactions:\\ the influence of hadronic spectrum and excluded volume effects on lattice thermodynamics and consequences on experiments
}
\author{P. Alba}
\affiliation{
Frankfurt Institute for Advanced Studies, Goethe Universit\"at Frankfurt,
D-60438 Frankfurt am Main, Germany}

\begin{abstract}
Repulsive hadronic interactions play a relevant role in the QCD dynamics, attractive ones being represented by resonance formation. In this study we propose different schemes in order to parameterise repulsive interactions, then being able to extract effective sizes of hadrons from fits to lattice QCD simulations. We find that allowing a difference between the strange and light sectors, strange particles are systematically smaller than light ones with equal mass. The very simple implementation of repulsive interactions would in principle allow to extract precise information about all hadronic species once corresponding lattice observables, sensitive to the species of interest, are provided. With the parameterisation which best reproduces lattice data there is also a good description of experimental yields measured by ALICE and STAR experiments.
\end{abstract}
 
\maketitle

\section{Introduction}

Nowadays lattice QCD simulations can provide very precise data on fluctuations of conserved charges  \cite{Bellwied:2015lba,Bazavov:2017dus}, which have been intensively studied as sensitive probes of the QCD transition \cite{Stephanov:1998dy}. Indeed, they should give clear signals for the presence of the Critical End Point (CEP), where sensitivity increases with the order of the fluctuation \cite{Stephanov:2008qz}. Recently higher moments of particle multiplicity distributions of net-proton \cite{Adamczyk:2013dal}, net-electric charge \cite{Adamczyk:2014fia} and net-kaon \cite{Adamczyk:2017wsl} have been measured by the STAR collaboration in the Beam Energy Scan, giving insights for a CEP around 14 GeV $\sqrt{s_{NN}}$, which corresponds to the high net-baryon density region of the QCD phase diagram.\\
Due to the sign problem it is not possible to study such a region on the lattice, which however can provide interesting informations at small chemical potential \cite{Bazavov:2017tot}, confirming no presence of the CEP at $\mu_B/T<2$ \cite{Bazavov:2017dus}. Effective models can circumvent this issue, and indeed recently two different approaches, which use respectively an holographic model \cite{Critelli:2017oub} and a vdW-HRG \cite{Vovchenko:2016rkn,Samanta:2017yhh}, have been able to predict the same location for the CEP ($\mu_B/T\simeq10$) once the same observables from lattice QCD at vanishing $\mu_B$ are used.\\
This could mean that a fingerprint of the CEP is already present at zero $\mu_B$ even for lower order observables, and this can be matched to models which employ a criticality, as it is done for example in the vdW-HRG by the balance of attractive and repulsive hadronic interactions. A similar result has been formerly obtained in \cite{Vovchenko:2016rkn}, where vdW parameters have been fixed to the liquid-phase transition in nuclear matter.\\
We study the balance between attractive and repulsive interactions, using extra higher mass resonances inspired by Quark-Model calculations for the first instead of a general attraction term for all particles as it is done in the vdW-HRG model; the physics behind the two approaches is different, but their effect on lattice observables is the same once repulsive interactions are considered.
Another important difference with respect to the vdW-HRG is that we do not consider point-like mesons, and we will show how even a small pion-radius can have relevant effects.\\
In addition we show a systematic study of repulsive interactions modelled by means of Excluded-Volume (EV) effects, in which we explore the specific effective hard-core sizes of hadrons depending on mass and quark content, allowing for example to a distinct behaviour between light and strange sectors.\\
The use of EV effects can be justified through the S-matrix approach, which correctly includes repulsive channels via experimentally measured phase shifts and gives results compatible with the hard-core approach \cite{Lo:2017ldt}. Using NN phase shifts, it has recently been shown how observables calculated on the lattice which are usually interpreted as a signal for deconfinement are indeed strictly connected to repulsive hadronic interactions \cite{Huovinen:2017ogf}. However the S-matrix approach is affected by large systematics due to the lack of experimental data on different elastic and inelastic interaction channels. On the other hand, EV-HRG allows to consistently account for all hadronic species and, once the radii are fixed, can give significant indications useful for the phase shifts approach and further inspire future experimental measurements at JLAB.\\
In the present paper we parameterise hadronic repulsive interactions by means of EV-HRG employing different particle lists and interaction schemes, thus extracting from lattice thermodynamics information on the effective sizes of hadrons, or namely their effective interactions.\\
This approach has already been successfully applied for analysing lattice simulations for SU2 and SU3 gauge theories \cite{Alba:2016fku}, with different schemes and accounting for higher mass states by means of an Hagedorn spectrum. Results showed a systematic presence of EV effects, with consistency between glueball masses in the two theories. Pure gauge is an exceptional benchmark since the particle content in the confined phase is clearer, while in the QCD case different flavours and quantum numbers play a role.\\

\section{The Hadron-Resonance Gas model}

Hadrons are the relevant degrees of freedom in the confined phase of QCD, and it is commonly accepted that this phase is well described by the Hadron-Resonance Gas (HRG) model up to the pseudo-critical temperature \cite{Borsanyi:2010bp}; however the crossover nature of the transition \cite{Aoki:2006we} does not allow to exactly identify a point in the QCD phase diagram where hadrons should completely disappear, and indeed studies on the spectral functions strongly suggest that hadrons progressively melt with increasing temperatures \cite{Burnier:2015tda,Burnier:2016kqm}. Furthermore, fit to experimental measurements of particle multiplicity distributions show quite a large uncertainty in the freeze-out temperature \cite{Cleymans:2005xv,Alba:2014eba}, with a maximum value of about 165 MeV for STAR measurements at the highest energy \cite{Adamczyk:2017iwn}.\\
The basic idea behind the HRG model is of describing a system of interacting hadrons as a gas of non-interacting hadrons and resonances, where resonance formation mediates the attractive interactions among the first \cite{Dashen:1969ep}. Thus it is possible to write the partition function as the sum of the independent contributions from all particles:
\begin{eqnarray}
\ln\mathcal{Z}(T,\{{\mu_B,\mu_Q,\mu_S}\})=\sum_{i \in Particles}(-1)^{{B_i}+1}\frac{{d_i}}{(2\pi^3)}\times\nonumber\\
\times\int d^3\vec{p}\,\ln\left[1+(-1)^{{B_i}+1}e^{-(\sqrt{\vec{p}^2+{m_i}^2}-{\mu_i})/T}\right]\,,\,\,\,
\label{partfunct}
\end{eqnarray}

where spin degeneracy $d_i$, mass $m_i$, baryon number $B_i$, electric charge $Q_i$, strangeness $S_i$ and single particle chemical potential $\mu_i=B_i\mu_B+Q_i\mu_Q+S_i\mu_S$ are used. Particle properties are usually taken from lists updated year by year \cite{Patrignani:2016xqp}.\\

From eq.\eqref{partfunct} fluctuations of conserved charges are defined as:
\begin{eqnarray}
\chi^{BQS}_{lmn}=\frac{\partial^{l+m+n}(\ln\mathcal Z/T^3)}{\partial(\mu_B/T)^l\partial(\mu_Q/T)^m\partial(\mu_S/T)^n}.
\end{eqnarray}

These are directly connected to the experiment, and it has been shown that the experimentally measured lower order moments are in agreement with the assumption of a thermalised hadronic medium \cite{Alba:2014eba,Alba:2015iva}.

\subsection{Excluded Volume effects}

Repulsive interactions can be implemented in the HRG model assuming that hadrons interact as hard spheres, thus giving an effective radius $r_i$ to particles \cite{Rischke:1991ke}. These interactions modify the thermodynamics, leading to a shifted single particle chemical potential given by:
\begin{equation}
\mu_i^*=\mu_i-v_i\,p\,\,,
\end{equation}
thus implying a transcendental equation for the pressure $p$, where:
\begin{equation}
v_i=\frac{16}{3}\pi\,r_i^3
\label{part_eigvol}
\end{equation}
is the particle eigenvolume.\\

All other observables are then obtained through thermodynamic relations, e.g. the net-baryon density is:
\begin{equation}
n_B(T,\vec{\mu})=\left(\frac{\partial p}{\partial\mu_B}\right)_T=\sum_i\frac{B_in_i^{id}(T,\mu_i^*)}{1+\sum_j v_j n_j^{id}(T,\mu_j^*)}\,\,\,.
\end{equation}

Naively speaking all intensive quantities are suppressed with respect to the ideal case, due to the extra volume introduced by the finite size of hadrons on top of the system volume. Similar results apply for other quantum numbers and for higher order fluctuations. In literature this version of the model is usually known as diagonal \emph{excluded volume} (EV-HRG) \cite{Vovchenko:2016ebv}.\\
Here it is possible to assign a different radius to every single particle, being so equally easy to parameterise the effective radii by quark content \cite{Andronic:2012ut,Alba:2016hwx}, mass \cite{Albright:2014gva,Alba:2016fku} and so on, allowing to easily separate the flavour dependence of interactions.\\
In this paper we explore possible differences between light and strange particles, as well as direct and inverse proportionality of eigenvolumes to hadron masses.\\
To our knowledge only the direct proportionality has been studied, being inspired by the bag model of hadrons \cite{Albright:2015uua,Vovchenko:2016ebv}, while the current poor knowledge about hadronic interactions in principle does not allow a clear understanding of the actual situation.\\
Obviously the EV-HRG is not the final answer to hadronic interactions, e.g. it can be seen that it is not consistent with the virial expansion of pressure already at second order; to do so one has to consider the proper interaction volume between the $ij$ particle pair through the following coefficients:
$$b_{ij}=\frac{2}{3}\pi\left(r_i+r_j\right)^3\,\,;$$
for $i=j$ one regains the values of $v_i$ in eq. \eqref{part_eigvol}, and in general it can be said that the proper inclusion of these \emph{crossterms} interactions (Cross-HRG) leads to a reduction in the magnitude of EV effects. The Cross-HRG in principle allows to properly treat any specific 2-body interaction, e.g. it is possible to account for particle-antiparticle annihilations which would further reduce EV magnitude \cite{Satarov:2016peb}.
The inclusion of the crossterms complicates the model leading to a set of coupled transcendental equations, one for each single particle pressure. For details see \cite{Vovchenko:2016ebv,Satarov:2016peb,Vovchenko:2017zpj}.\\
Currently there are different studies on improved versions of the EV-HRG \cite{Vovchenko:2017cbu}, but in the following sections we will concentrate on the diagonal version of the EV which is able to catch all the physics of interest between light and strange sectors. It is however worth to note that the Cross-HRG gives the same qualitative behaviour for most of the observables available from lattice, with no significant changes in fit results.\\

\section{Particle lists}

The list of particles plays a major role in the thermodynamics of the HRG, with the higher mass resonances being more influential in the high temperature range. Albeit for common thermodynamic observables (pressure, energy density, etc.) the inclusion of more states straightforwardly increases their values due to the inclusion of more \emph{attraction} in the system, higher order fluctuations can extract selective information from different sectors of the hadronic spectrum. Indeed, from very precise lattice simulations it is possible to construct combinations of fluctuations which are sensitive to specific set of hadronic quantum numbers \cite{Bazavov:2014xya,Alba:2017mqu}; this has shown how the standard list of measured hadronic states is not suitable for a coherent description of all lattice results and that there is still the need for more states, and/or new physics, being them awaiting for confirmation by the Particle Data Group (PDG) \cite{Patrignani:2016xqp} or calculated from the Quark Model (QM) \cite{Ebert:2009ub,Capstick:1986bm}.\\
We employ different versions of the PDG list (2005, 2014, 2016), and a list in which QM states are used. This way due to the intrinsic differences among lists, it is possible to better track the importance of different hadronic sectors, in particular the one of strange baryons. The content of the different lists is summarised in table \ref{list_part_content}.\\
PDG2005 consists of old and very well established states, while PDG2014 and PDG2016 versions are improved lists, with essentially the same light content and a modest difference in the strange sector, which anyway will be relevant for strange baryon observables.\\
The effect of extra higher mass states on particle yields will be studied in \cite{mio_forthcoming}, where it is clear how these play a crucial role in the description of these quantities.\\
We include the $\sigma$ meson for PDG2014 and QM lists. It has been shown how the repulsive interactions deriving from phase shifts data in the ($\pi$-$\pi$,\,\,I=2) channel counterbalance the attraction due to the presence of this meson \cite{Broniowski:2015oha}, but since in our approach all repulsive interactions are already accounted through effective hadronic sizes we think it is more consistent to include the $\sigma$. A similar argument would apply to the $\kappa$ strange meson, which however is not include in the present study since its existence is currently still highly debated.\\

\begin{table}[h]
\begin{tabular}{|c|c|c|c|c|c|c|c|c|c|}
\hline
 & $\pi^0, ...$ & $\pi^+, ...$ & $K^+$ & $p, N^+$ & $\Delta^{++}$ & $\Lambda$ & $\Sigma^0$ & $\Xi^-$ & $\Omega^-$\\
\hline
PDG05 & 24 & 8 & 7 & 5 & 3 & 4 & 2 & 2 & 1 \\
\hline
PDG14 & 63 & 22 & 16 & 27 & 14 & 18 & 20 & 6 & 2 \\
\hline
PDG16 & 78 & 29 & 23 & 28 & 22 & 19 & 22 & 11 & 4 \\
\hline
QM & 202 & 64 & 42 & 48 & 27 & 48 & 51 & 47 & 15 \\
\hline
\end{tabular}
\caption{Particle content for the different lists. Columns from left to right read as follow (with total multiplicity accounting for anti-particles and isospin degeneracy in parenthesis): uncharged light mesons (1), charged light mesons (2), charged kaons (4), charged nucleons (4), $\Delta$ particles (8), $\Lambda$ baryons (2), $\Sigma$ baryons (6), $\Xi$ baryons (4), $\Omega$ baryons (2).}
\label{list_part_content}
\end{table}

\section{Fit to lattice thermodynamics}
We perform fits to observables calculated via lattice simulations in order to extract properties on the effective radii of hadrons and resonances.\\
This is definitively interesting, since the current knowledge on hadron sizes is quite poor. Actually only the charge radii of a few ground states have been experimentally measured, see table \ref{tab_charged_radii}. The very few available experimental data do not allow for a conclusive argument on any trivial trend in hadronic sizes, but it can be guessed that strange states, even with a larger mass, have smaller sizes with respect to light ones.\\
\begin{table}[h]
\begin{tabular}{|c|c|c|c|c|c|}
\hline
(fm) & $\pi^{\pm}$& $K^{\pm}$& p & $\Sigma^{-}$\\
\hline
$\sqrt{\langle r_E^2\rangle}$ &0.672$\pm$0.008 & 0.569$\pm$0.031 & 0.8751$\pm$0.0061 & 0.78$\pm$0.10 \\
\hline
$\sqrt{\langle r_M^2\rangle}$ & \textbackslash\textbackslash & \textbackslash\textbackslash & 0.78$\pm$0.04 & \textbackslash\textbackslash\\
\hline
\end{tabular}
\caption{
Experimental estimates of charge radii, electric and magnetic, for different ground states \cite{Patrignani:2016xqp}.
}
\label{tab_charged_radii}
\end{table}

\begin{figure*}[htbp]
\begin{center}
\includegraphics[width=0.3\textwidth]{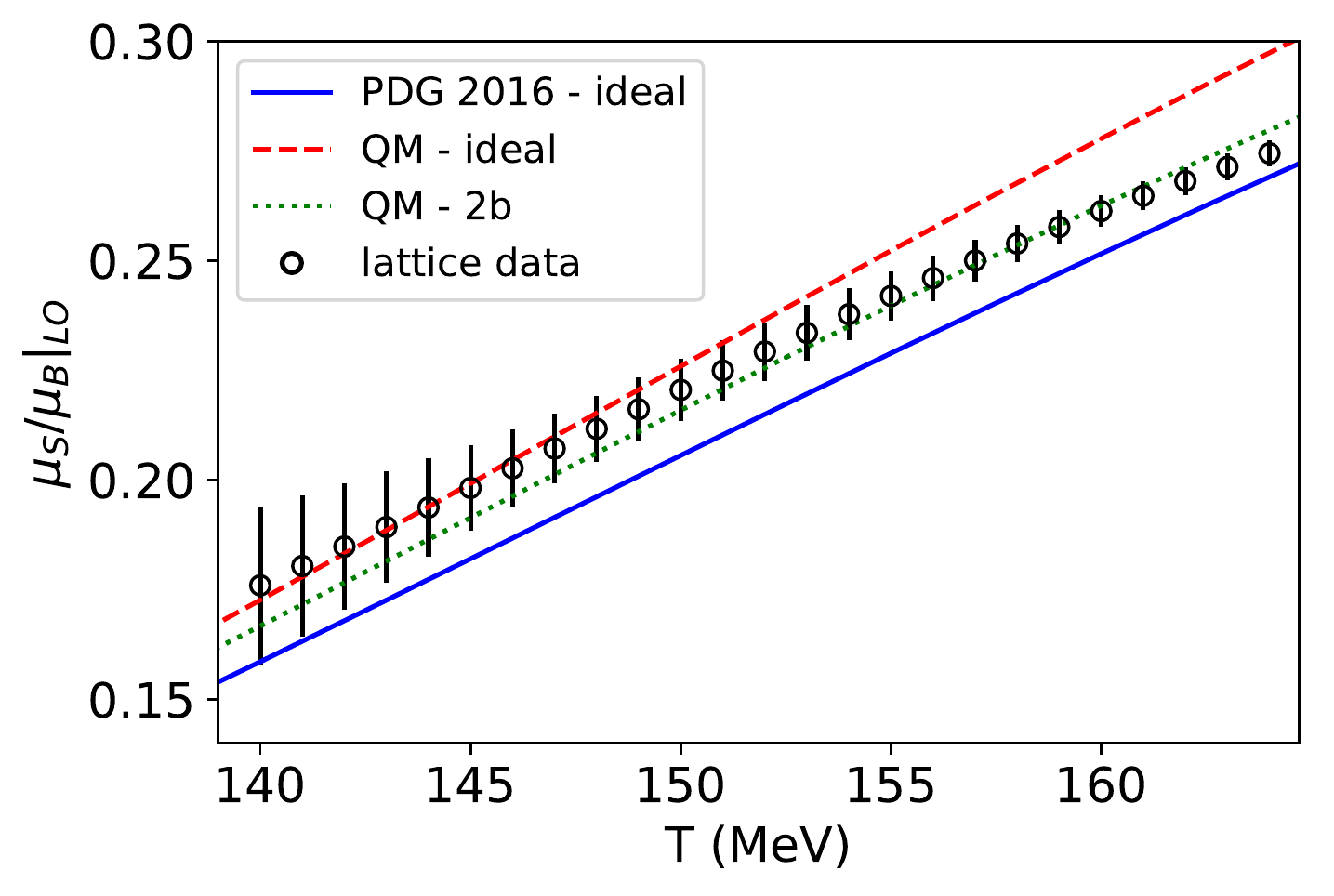}~~\includegraphics[width=0.3\textwidth]{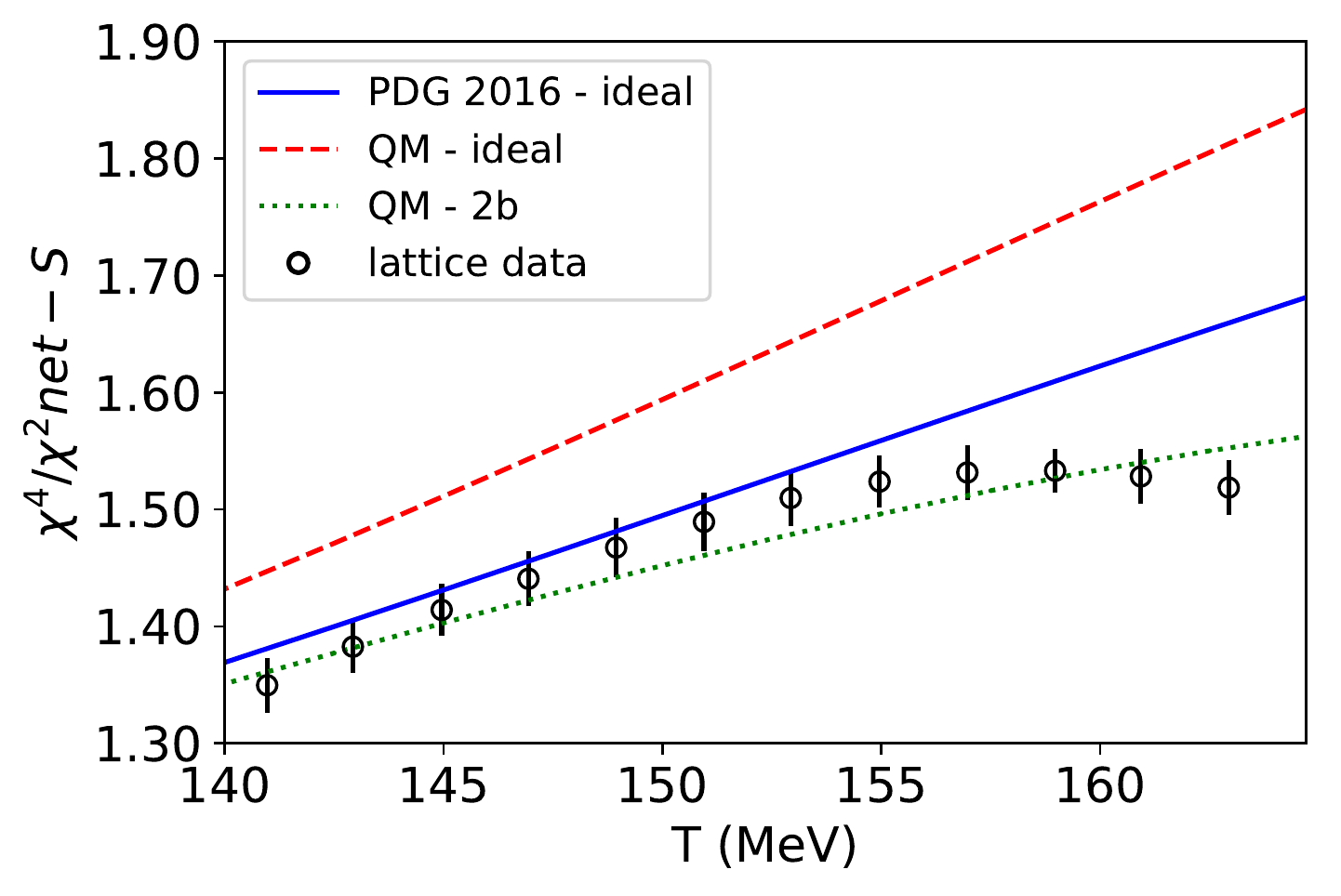}~~\includegraphics[width=0.3\textwidth]{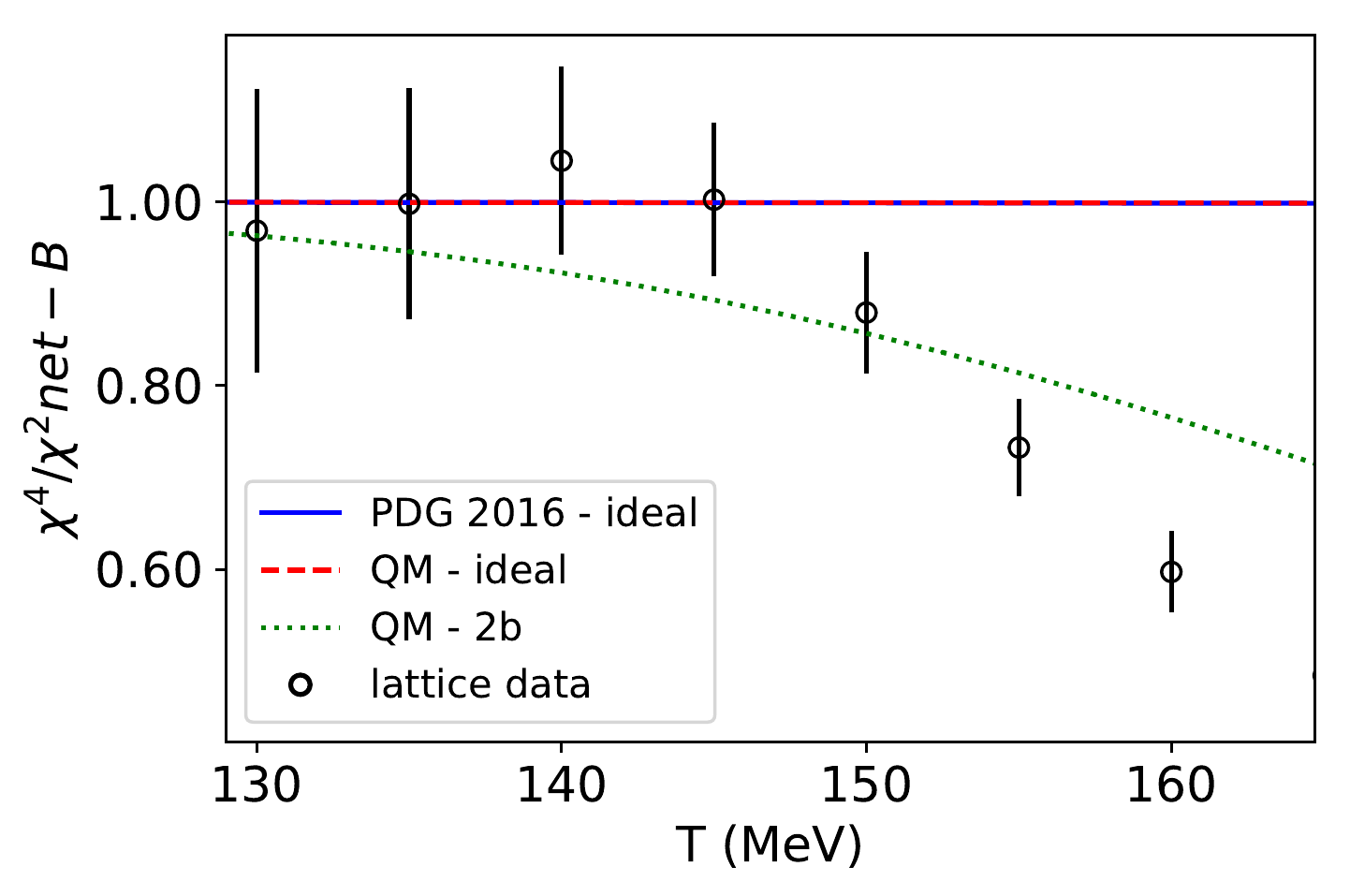}
\caption{Color online: lattice data for $\mu_S/\mu_B|_{LO}$ \cite{Borsanyi:2013hza}, $\chi_4^S/\chi_2^S$ \cite{Bellwied:2013cta} and $\chi_4^B/\chi_2^B$ \cite{Bazavov:2017dus} in comparison to HRG calculations with PDG2016 and QM lists in the ideal scheme (blue continue and red dashed curves) and with QM list in the EV-HRG with the corresponding parameterisation listed in table \ref{fit-2b} (green dotted curves).}
\label{plot-linee}
\end{center}
\end{figure*}

Keeping all of this in mind, we perform a systematic study on differences between light and strange sectors, allowing resonances to have very different behaviours with respect to ground states; namely we consider different combinations of the following schemes: fixed radii (\emph{r}) for all particles, radii directly (\emph{b}) and inversely (\emph{inv}) proportional to particle mass. For the sake of simplicity, we will parameterise different EV schemes by means of ground state radii ($\pi$, $K$, $p$ and $\Lambda$) \cite{Alba:2016fku}, which will be specified time by time.\\
We use data from lattice simulations, extrapolated to the continuum with physical values for the quark masses, for the following observables: pressure, interaction measure \cite{Borsanyi:2013bia}, $\chi_{11}^{ud}$, $\chi_{11}^{us}$ \cite{Bellwied:2015lba}, $\chi_{11}^{ss}$, $\chi_4^l/\chi_2^l$, $\chi_4^S/\chi_2^S$ \cite{Bellwied:2013cta}, $\chi_4^B/\chi_2^B$ \cite{Bazavov:2017dus} and $\mu_S/\mu_B|_{LO}$ \cite{Borsanyi:2013hza}. We restrict our study in the temperature range between 110 and 164 MeV, for about 100 lattice points; it should be noted that while there is no real reason to fix a lower bound in temperature, if not due to availability of lattice simulations, we choose such an upper value inspired by the current estimate for the pseudo-critical temperature and for the chemical freeze-out one.
Anyhow a smaller upper bound of 160 MeV leads to a tiny difference in the total number of lattice points, with no modifications in the results of fits.\\
We perform the fits minimising the $\chi^2$ defined in the following way:
\begin{equation}
\label{chiquadro}
\chi ^2~=~\frac{1}{N_{\rm dof}}\sum_{h=1}^N \frac{\left( \langle x_h^{\rm latt}\rangle~-~
\langle x_h \rangle \right)^2}{\sigma_h^2}~,
\end{equation}
where $\langle x_h^{\rm latt}\rangle $ and $\langle x_h\rangle $ are respectively the values obtained from lattice and HRG for a specific observable at a specific temperature, $\sigma_h$ is the corresponding uncertainty from lattice, and $N_{\rm dof}$ is the number of degrees of freedom, i.e. the number of data points $N$ minus the number of fitting parameters.
Uncertainties on fitted parameters are obtained through the $\chi^2+1$ criterium.\\ 
As a crosscheck we performed the fits using as an estimator the average of the $\chi^2$s for single observables, in order to equally weight observables with a different number of points in the chosen temperature range, finding that there is no modification in final conclusions.\\
As a final remark, we have chosen the set of data in order to have as few correlation as possible among different observables and selecting the most relevant physical differences between strange and light sectors.\\

\section{Results from lattice fit}
In table \ref{fit_ideal} are listed the $\chi^2$ values obtained without any EV effect (\emph{ideal} case). It is clear how the very small PDG2005 gives a poor description of the available lattice data, while PDG2014 and PDG2016 significantly improve the prediction power of the HRG model.\\
\begin{table}[h]
\begin{tabular}{|c|c|c|c|c|c|}
\hline
 & PDG05& PDG14& PDG16& QM\\
\hline
$\chi^2$ & 49.645 & 10.094 & 9.331 & 16.312\\
\hline
\end{tabular}
\caption{$\chi^2$ obtained from different particle lists with no EV effects.}
\label{fit_ideal}
\end{table}

On the basis of the $\mu_S/\mu_B|_{LO}$(\footnote{this quantity is proportional to $\chi_{11}^{BS}/\chi_2^S$.}) it has been argued that PDG lists are missing strange baryons \cite{Bazavov:2014xya}. This gap can be filled by QM states, with however a consequent worsening of the $\chi_4^S/\chi_2^S$ due to the competitive effect of multi-strange baryons(\footnote{this quantity is proportional to the averaged squared net-strangeness $\langle S^2\rangle$.}) (see figure \ref{plot-linee}). The net effect is a larger $\chi^2$ for the QM with respect to PDG2014 and PDG2016.\\
S=1 particles could bring the $\chi_4^S/\chi_2^S$ down, essentially counteracting the effect of multi-strange baryons, but from this point of view the QM gives already all possible states from quark combinations. So to push the agreement with the lattice one should try to find the best criterium in order to select hadrons which enter the particle list \cite{Alba:2017mqu}, or should rely on new physics like the one given by repulsive interactions.\\
\begin{figure}[htbp]
\begin{center}
\includegraphics[width=0.24\textwidth]{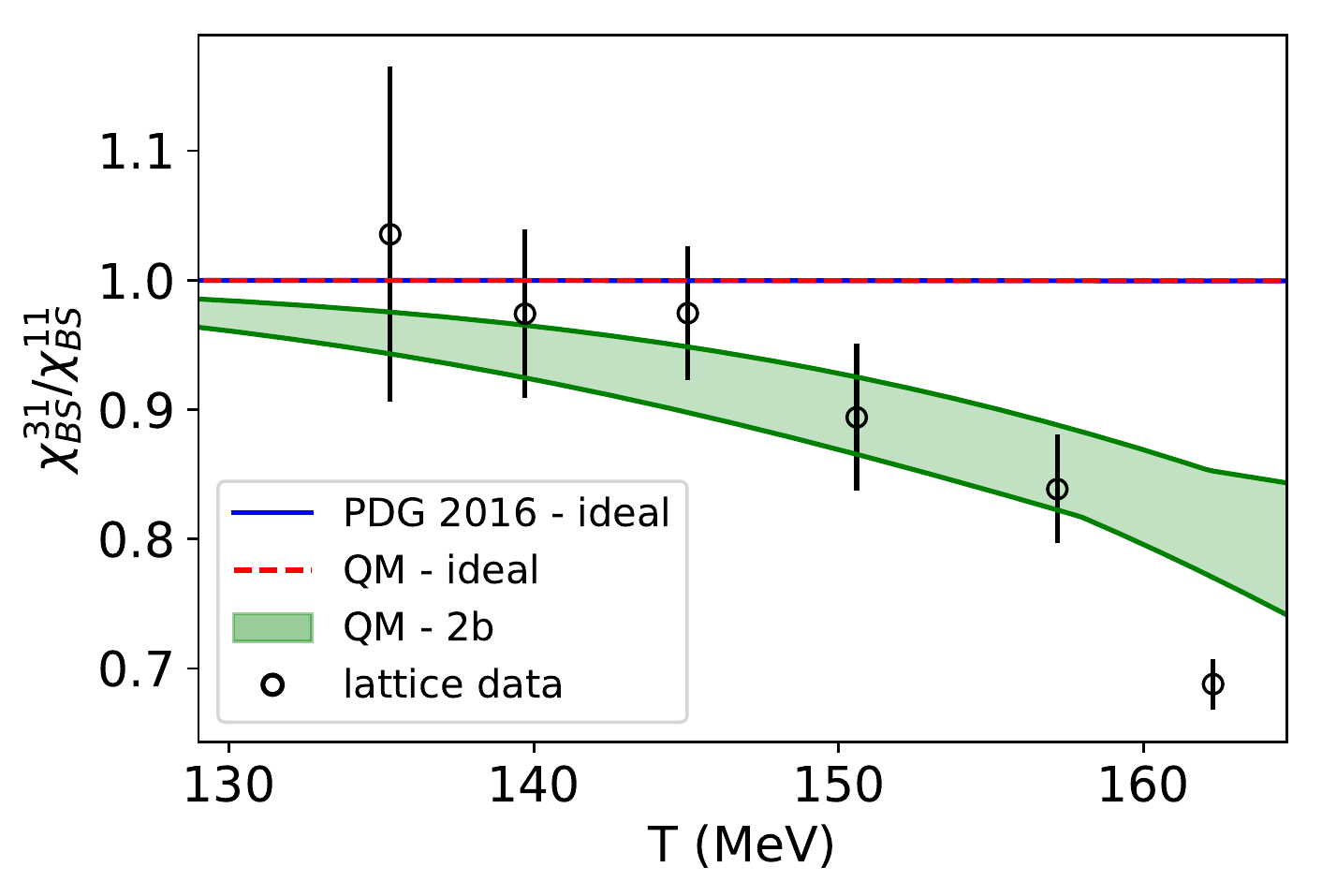}~~\includegraphics[width=0.24\textwidth]{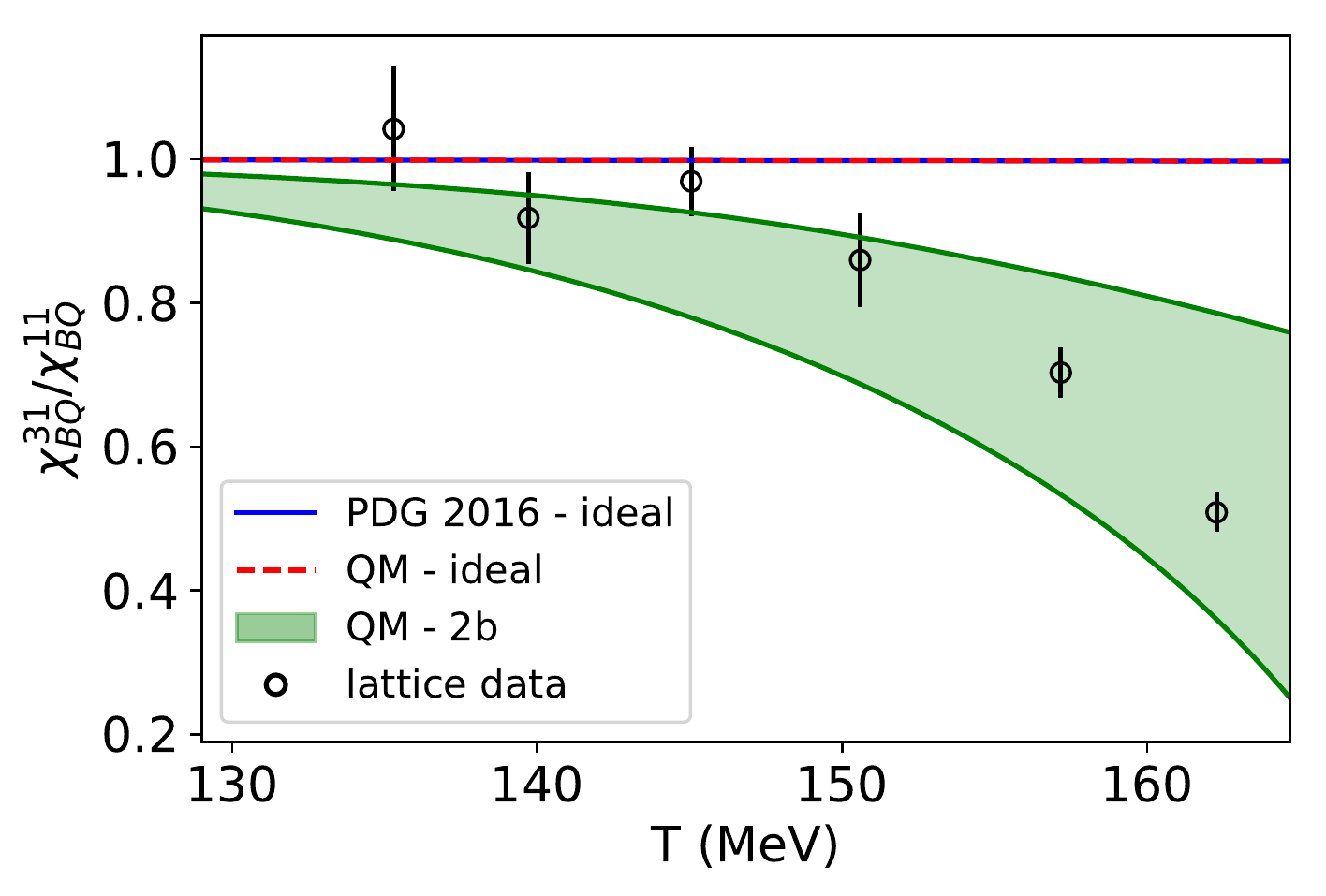}\\
\includegraphics[width=0.24\textwidth]{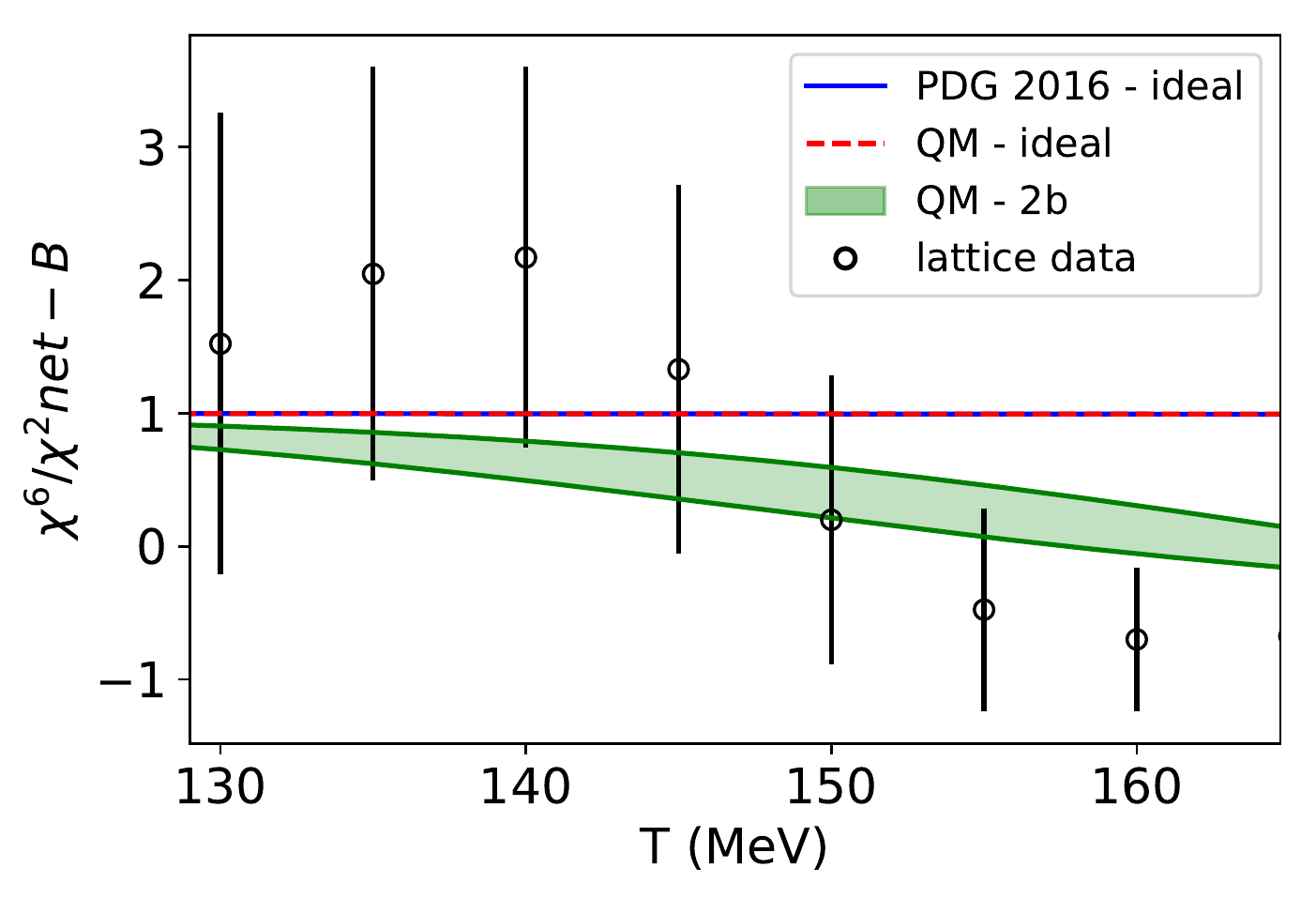}~~\includegraphics[width=0.24\textwidth]{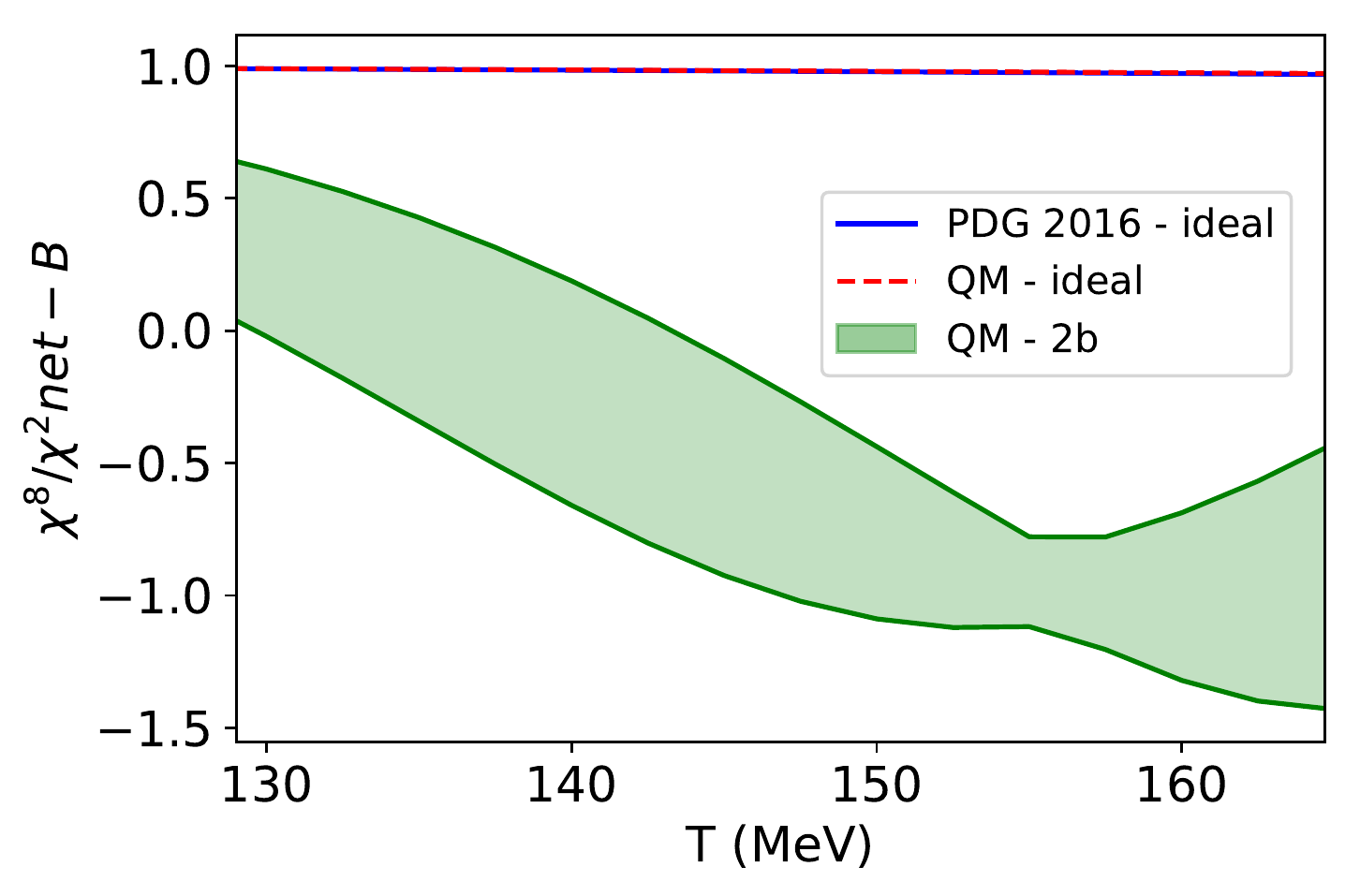}
\caption{Color online: lattice data for $\chi_{31}^{BS}/\chi_{11}^{BS}$, $\chi_{31}^{BQ}/\chi_{11}^{BQ}$ and $\chi_6^B/\chi_2^B$ \cite{Karsch:2017mvg,Karsch:2016yzt,Bazavov:2017dus} in comparison to HRG calculations with PDG2016 and QM lists in the ideal scheme (blue continue and red dashed lines) and with QM list plus the corresponding EV effects listed in table \ref{fit-2b} with errors (green dashed area). Predictions for $\chi_8^B/\chi_2^B$ are shown. The lattice data here shown are not considered for the fits.}
\label{plot-bande}
\end{center}
\end{figure}

\begin{figure*}[htbp]
\begin{center}
\includegraphics[width=0.4\textwidth]{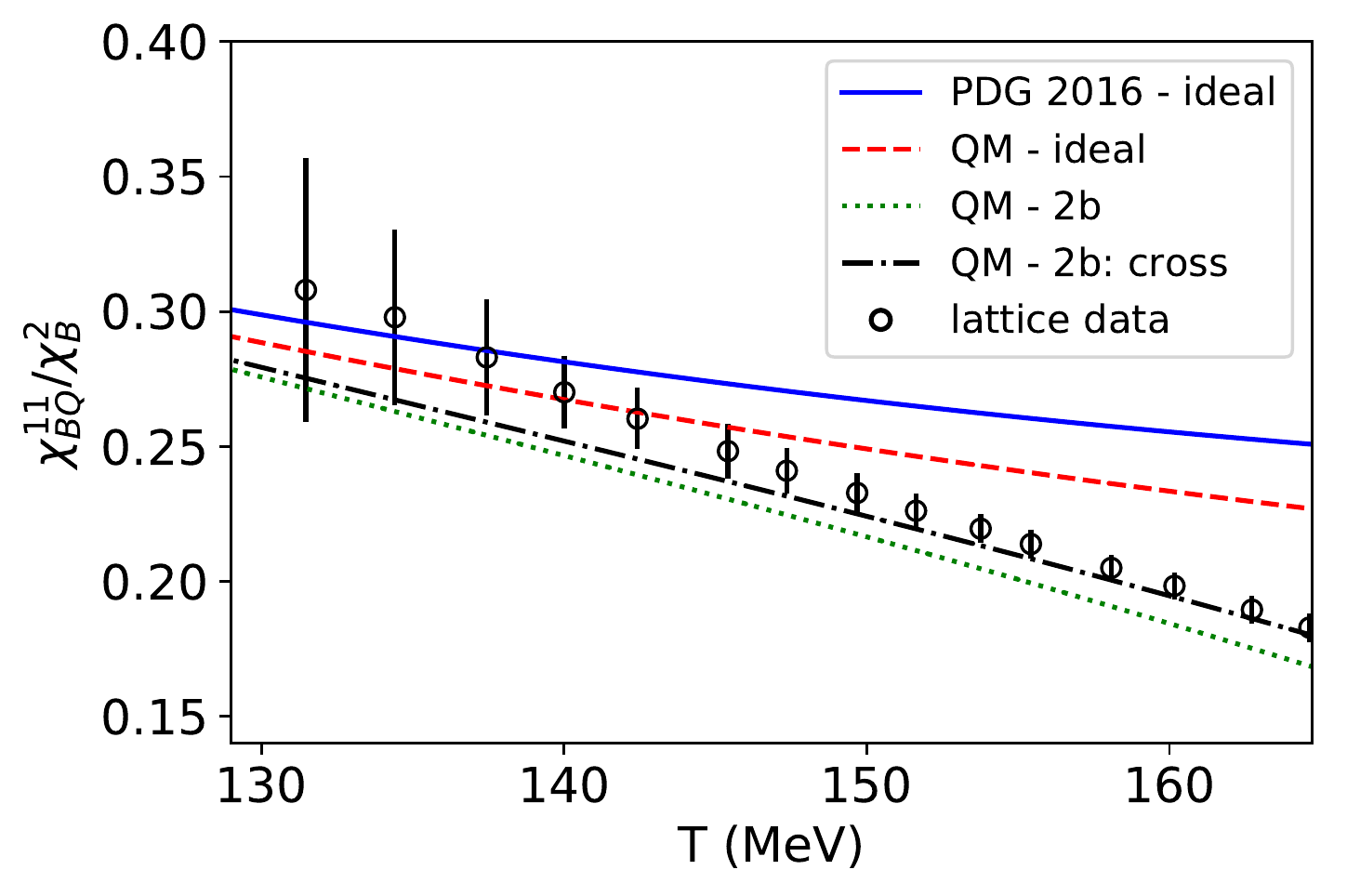}~~\includegraphics[width=0.4\textwidth]{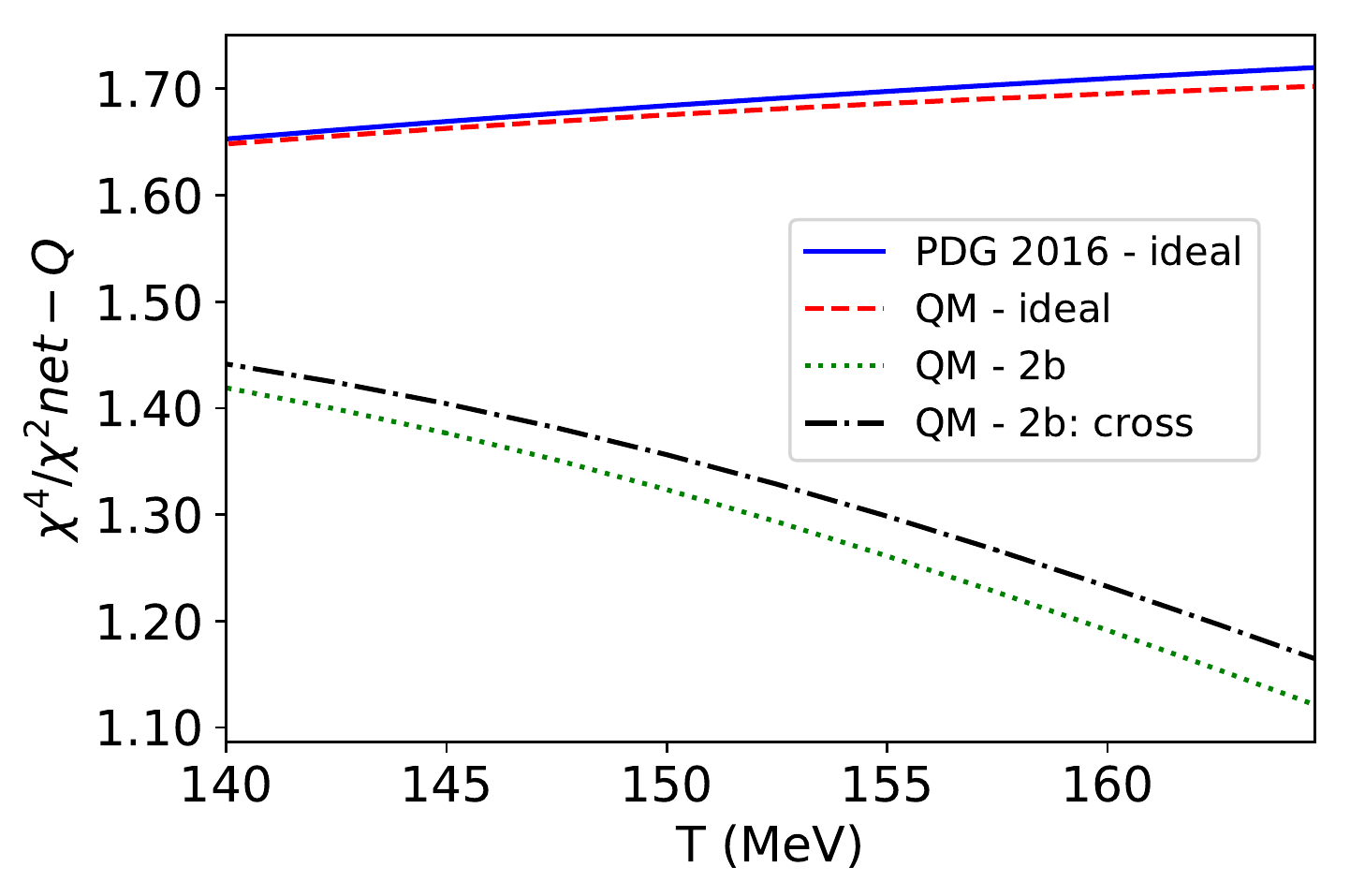}
\caption{Color online: lattice data for $\chi_{11}^{BQ}/\chi_{2}^B$ \cite{Karsch:2017zzw,Bazavov:2017dus} in comparison to HRG calculations with PDG2016 and QM lists in the ideal scheme (blue continue and red dashed curves) and with QM list in the EV-HRG and Cross-HRG with the corresponding parameterisation listed in table \ref{fit-2b} (green dotted and black dot-dashed curves). Predictions for $\chi_4^Q/\chi_2^Q$ are shown. The lattice data here shown are not considered for the fits.}
\label{plot-cross}
\end{center}
\end{figure*}
In table \ref{1-par} results of fits performed with only one parameter are listed. The introduction of EV effects generally improves lattice description, with the non trivial result of a finite proton radius $r_p$. In this case the best $\chi^2$ is given by the combined use of QM list and an eigenvolume which increases with hadronic mass. This is similar to what was found in \cite{Alba:2016fku} for the pure gauge. However the different flavours and quantum numbers present in QCD allow for a deeper study.\\
\begin{table}[h]
\begin{tabular}{|c|c|c|c|c|c|}
\hline
 & r & b& inv\\
\hline
PDG05 & $\chi^2$=49.33 & $\chi^2$=49.645 & $\chi^2$=35.215 \\
 & $r_p$=0.152$\pm$0.095 & $r_p$=0.007$\pm$0.327 & $r_p$=0.419$\pm$0.051 \\
\hline
PDG14 & $\chi^2$=9.248 & $\chi^2$=9.848 & $\chi^2$=9.062 \\
 & $r_p$=0.174$\pm$0.068 & $r_p$=0.13$\pm$0.089 & $r_p$=0.162$\pm$0.067 \\
\hline
PDG16 & $\chi^2$=6.814 & $\chi^2$=7.549 & $\chi^2$=7.883 \\
 & $r_p$=0.211$\pm$0.051 & $r_p$=0.181$\pm$0.049 & $r_p$=0.17$\pm$0.061 \\
\hline
QM & $\chi^2$=6.945 & $\chi^2$=3.784 & $\chi^2$=15.097 \\
 & $r_p$=0.269$\pm$0.038 & $r_p$=0.249$\pm$0.028 & $r_p$=0.151$\pm$0.055 \\
\hline
\end{tabular}
\caption{Proton radius obtained from a fit to lattice data in the r, b and inv schemes and different particle lists.}
\label{1-par}
\end{table}

In tables \ref{fit-2r},\ref{fit-2b} and \ref{fit-sinv} we show the results for fits obtained considering respectively: one radius for all light particles and one for strange ones (\emph{2r} scheme), radii increasing with particle mass but with different proportionality constants for light and strange particles (\emph{2b} scheme), and lastly the same as before but with strange radii decreasing with particle mass (\emph{s-inv} scheme). Even if the last scheme could sound odd and counterintuitive, it has been found to be relevant for fit to particle yields \cite{Alba:2016hwx}.\\
The introduction of one additional parameter generally improves the quality of the fit, but the most interesting result that can be drawn is that, irrespective of the scheme employed, strange particles have a systematically smaller radii than the corresponding light ones with equal mass\footnote{in the s-inv scheme this is true only for baryons.}.\\
As already pointed out the actual knowledge on the sizes of hadrons and resonances is rather poor, but our finding can be supported by several interconnected arguments. Indeed phenomenological cross sections for strange particles are smaller than light ones, which can be naively connected to a smaller effective interaction area; further speaking this is what one would expect from the Quark Model: strange quarks, being heavier than the u-d ones, result into more localised bound states with reduced radial excitations and angular momenta.\\
Taking into consideration the different strange-baryon content of the lists under investigation, it should be noted how the $\Lambda$ radius $r_\Lambda$ in table \ref{fit-2b} evolves from being zero for PDG2005 to a finite value of 0.266 fm for the QM list with rather small errors, with a consequent gradual improvement of the $\chi^2$. The same is confirmed by fits with more parameters, with however any critical improvement in the quality of the fit, or conversely with no improvement in the corresponding p-values.\\

\begin{table}[h]
\begin{tabular}{|c|c|c|c|c|c|}
\hline
 & $\chi^2$& $r_p$ (fm)& $r_\Lambda$ (fm)\\
\hline
PDG05 &44.3 & 0.446 $\pm$0.115 & 0.173 $\pm$0.133 \\
\hline
PDG14 &5.723 & 0.389 $\pm$0.101 & 0.173 $\pm$0.1 \\
\hline
PDG16 & 4.28 & 0.383 $\pm$0.1 & 0.217 $\pm$0.066 \\
\hline
QM & 6.263 & 0.351 $\pm$0.099 & 0.274 $\pm$0.044 \\
\hline
\end{tabular}
\caption{Proton and $\Lambda$ radii obtained from the fit to lattice data in the 2r scheme and different particle lists.}
\label{fit-2r}
\end{table}

\begin{table}[h]
\begin{tabular}{|c|c|c|c|c|c|}
\hline
 & $\chi^2$& $r_p$ (fm)& $r_\Lambda$ (fm)\\
\hline
PDG05 & 45.48 & 0.394 $\pm$0.093 & 0.004 $\pm$0.432 \\
\hline
PDG14 & 4.719 & 0.375 $\pm$0.081 & 0.016 $\pm$0.508 \\
\hline
PDG16 & 3.595 & 0.373 $\pm$0.085 & 0.172 $\pm$0.073 \\
\hline
QM & 1.714 & 0.38 $\pm$0.092 & 0.266 $\pm$0.034 \\
\hline
\end{tabular}
\caption{Proton and $\Lambda$ radii obtained from the fit to lattice data in the 2b scheme and different particle lists.}
\label{fit-2b}
\end{table}

\begin{table}[h]
\begin{tabular}{|c|c|c|c|c|c|}
\hline
 & $\chi^2$& $r_p$ (fm)& $r_\Lambda$ (fm)\\
\hline
PDG05 & 40.632 & 0.487 $\pm$0.157 & 0.249 $\pm$0.052 \\
\hline
PDG14 & 3.717 & 0.404 $\pm$0.099 & 0.171 $\pm$0.063 \\
\hline
PDG16 & 2.26 & 0.391 $\pm$0.092 & 0.192 $\pm$0.051 \\
\hline
QM & 8.585 & 0.353 $\pm$0.078 & 0.201 $\pm$0.043 \\
\hline
\end{tabular}
\caption{Proton and $\Lambda$ radii obtained from the fit to lattice data in the s-inv scheme and different particle lists.}
\label{fit-sinv}
\end{table}

\subsection{Observables in the best scenario}
The best $\chi^2$ is given by the combined use of the QM list and a 2b scheme for light and strange particles. In the following we will compare the results obtained with the corresponding parameterisation of table \ref{fit-2b} with respect to PDG2016 and QM lists in the ideal case and to lattice data.\\
In figure \ref{plot-linee} it is shown how EV effects have a modest influence on the $\mu_S/\mu_B|_{LO}$, slightly improving the HRG result at higher temperatures, while they are responsible for a suppression in the $\chi_4^S/\chi_2^S$ which provides a final result comparable to the PDG2016 list in the ideal case. A similar suppression can be seen in different observables, e.g. the $\chi_4^B/\chi_2^B$ (figure \ref{plot-linee}), and in the $\chi_{31}^{BQ}/\chi_2^B$ and $\chi_{31}^{BS}/\chi_2^B$ (upper panels figure \ref{plot-bande}). In general 4th order derivatives, diagonal and non, show with respect to the 2nd order ones a systematic difference which is compatible to the differential suppression due to EV effects, which however extends to higher order fluctuations. Indeed for $\chi_6^B/\chi_2^B$ and $\chi_8^B/\chi_2^B$ lattice simulations predict non-monotonic behaviours, in particular a change of sign at higher temperatures, which result compatible with EV effects (see lower panels of figure \ref{plot-bande}).\\
It is generally clear how the standard HRG is not able to reproduce any of these aspects even for temperatures which should be compatible with the hadronic phase, while this is a natural result of repulsive interactions.\\
In figure \ref{plot-cross}, left panel, results for the $\chi_{11}^{BQ}/\chi_2^B$ are shown; the ideal HRG fails in describing such a quantity already at 145 MeV, while EV effects naturally bring the result into an agreement which is further improved by Cross-HRG, with no changes in the employed parameterisation.\\
Indeed observables connected to net-electric charge are the most sensitive to changes in the system; another example is given by the $\chi_4^Q/\chi_2^Q$, shown in the right panel of figure \ref{plot-cross}, which can give a clear signal of EV effects since it is mostly influenced by lighter charged particles as pions, regardless to other higher mass particles. This can be extremely interesting, since the suppression here found is not present in the vdW-HRG \cite{vold_privatecomm}, where all mesons do not interact at all. Future lattice calculations of this observable could confirm the presence of mesonic interactions if the corresponding suppression of 4th to 2nd ratios is seen; furthermore this quantity can be directly compared to measurements of net-electric charge multiplicity distribution of heavy-ion collision at LHC, which correspond to the $\mu_B\simeq0$ region of the QCD phase diagram.

\section{Fit to particle yields}

One of the main achievements of the statistical model is the description of particle production in heavy-ion collisions by means of few parameters, namely temperature $T$, baryon chemical potential $\mu_B$ and system volume per unit of rapidity $V$, all evaluated at chemical freeze-out (see, e.g., Refs.~\cite{Rafelski:1982pu,Cleymans:1992zc,BraunMunzinger:1994xr,Becattini:2000jw,BraunMunzinger:2001ip,Becattini:2005xt,Andronic:2008gu,Andronic:2016nof}).\\
Initial conditions, deriving from colliding nuclei, are implemented in the model through effective $\mu_S$ and $\mu_Q$. Their $T$ and $\mu_B$ dependences are obtained imposing strangeness neutrality $N_S=0$ and isospin imbalance $N_Q/N_B=0.4$ (where 0.4 is a typical value valid for most of AA collisions).\\
Final particle yields are obtained adding to the primordial thermal yield the contribution from resonances, which is given by:
\begin{equation}
\label{Nh}
\langle N_h \rangle ~=~V\,n_h~+~V\,\sum_R\langle n_h\rangle_R\,n_R~,
\end{equation}
where $\langle n_h\rangle_R$ is the average number of particles of type $h$ resulting from a
decay of resonance $R$, and $n_i$ is the thermal density calculated through the statistical model \cite{Nahrgang:2014fza}. For a detailed description of the procedure employed in modelling QM decays see \cite{mio_forthcoming}.\\

\begin{figure*}[htbp]
\begin{center}
\includegraphics[width=0.5\textwidth]{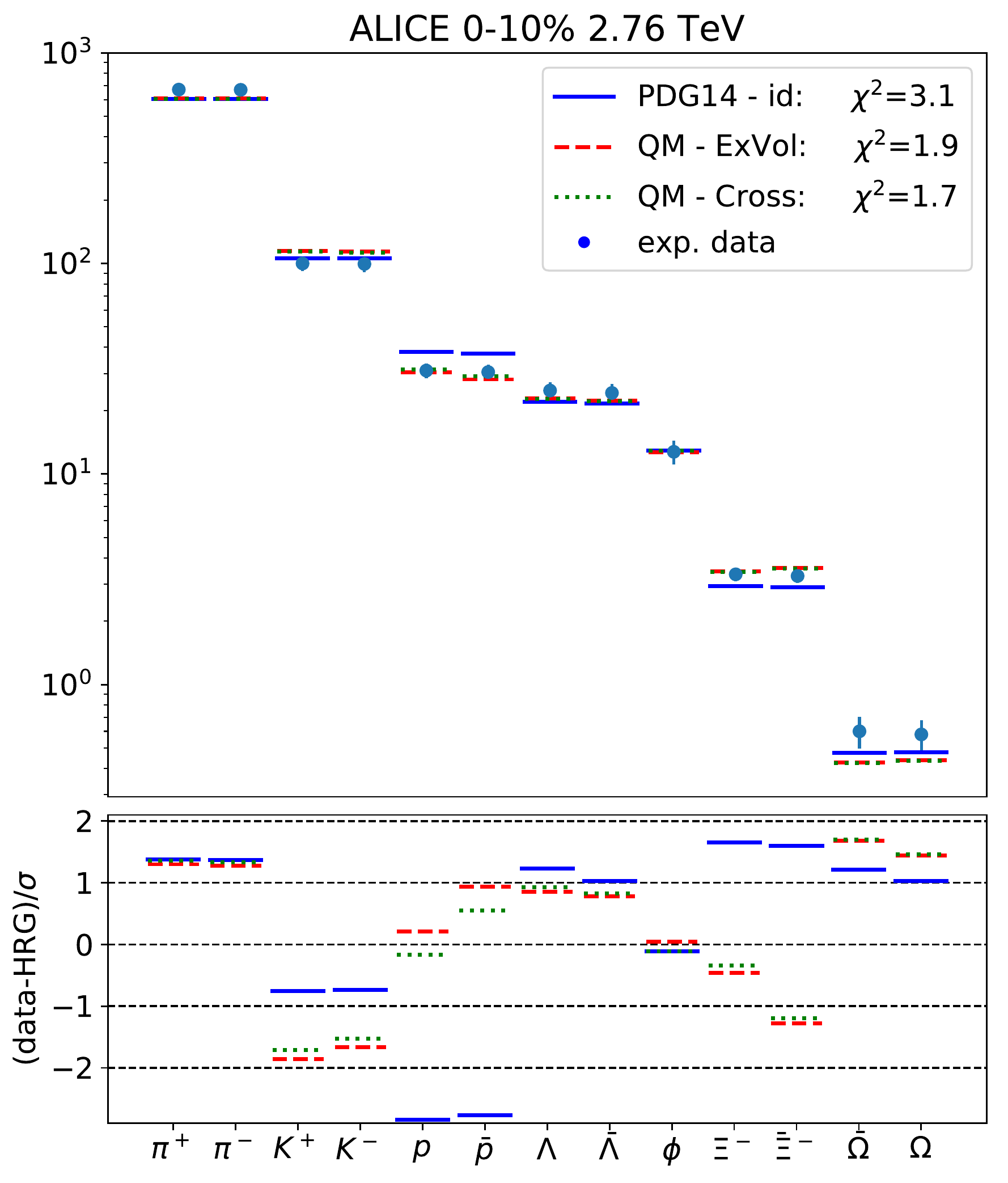}~~\includegraphics[width=0.5\textwidth]{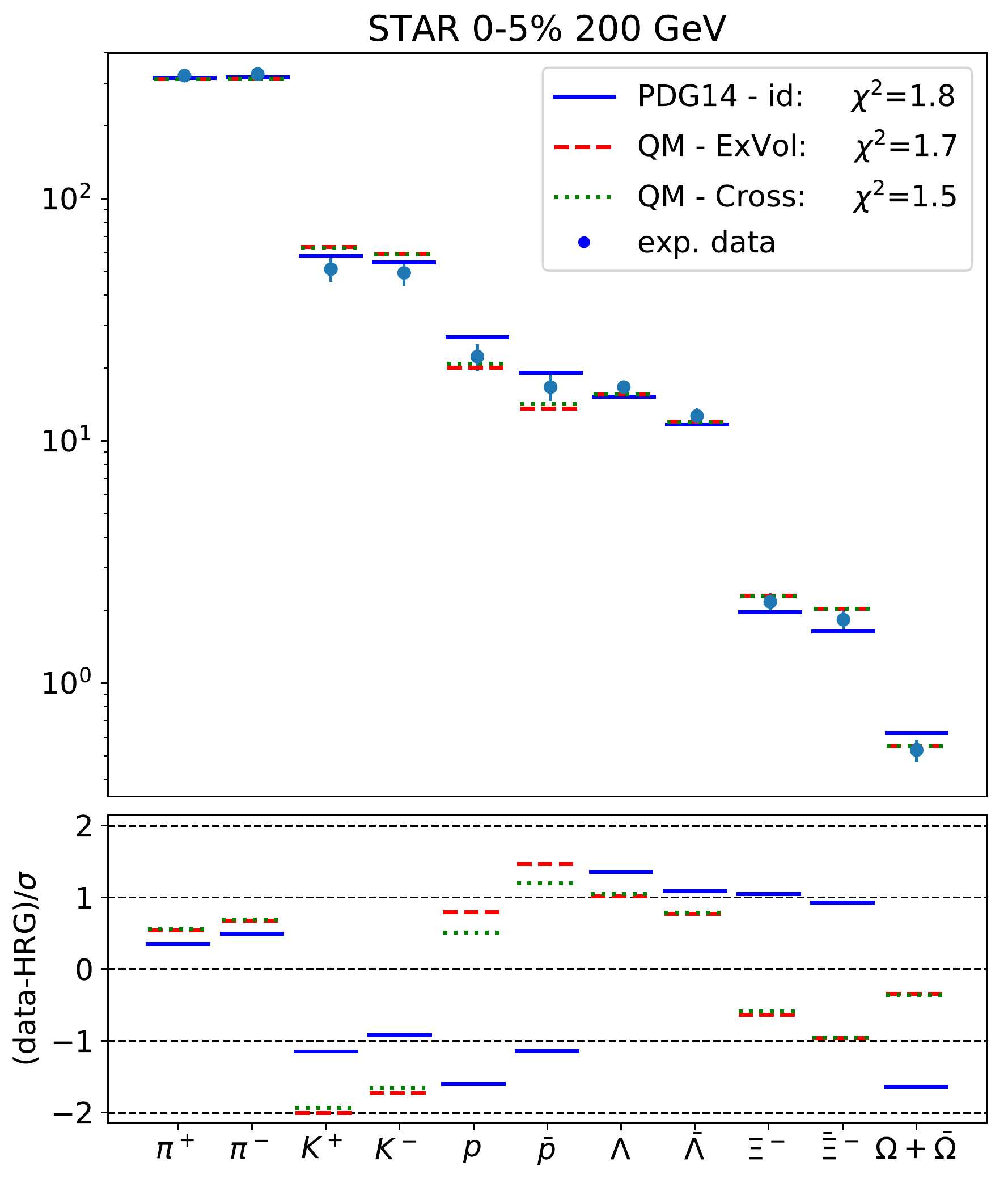}
\caption{Color online: yields of hadrons produced in Pb-Pb collisions measured by ALICE and STAR in comparison to results from ideal HRG, EV-HRG and Cross-HRG fits with PDG2014 and QM lists and parameters extracted from lattice which are listed in table \ref{fit-2b}. Deviations in units of experimental uncertainties are shown in lower panels.}
\label{alice-yields}
\end{center}
\end{figure*}

For the fit we minimise the $\chi^2$ defined in eq. \eqref{chiquadro} where are used, instead of lattice points, data for particle yields at mid-rapidity measured respectively by ALICE at 2.76 TeV and by STAR at 200 GeV; in detail we use:
\begin{description}
\item[ALICE] $\pi^{\pm}$, $K^{\pm}$, $p$($\bar p$) \cite{Abelev:2013vea}, $\Lambda$($\bar\Lambda$) \cite{Abelev:2013xaa,Schuchmann:2015lay} , $\Xi^{\pm}$, $\Omega^{\pm}$ \cite{ABELEV:2013zaa} , $\phi$ \cite{Abelev:2014uua} at 0-10\% centrality;
\item[STAR] $\pi^{\pm}$, $K^{\pm}$, $p$($\bar p$) \cite{Adams:2003xp}, $\Lambda$($\bar\Lambda$), $\Xi^{\pm}$, $\Omega^{-}+\Omega^{+}$ \cite{Adams:2006ke} at 0-5\% centrality.
\end{description}

In figure \ref{alice-yields} are shown the results of fits obtained with PDG2014 using the ideal HRG, and with the QM list using EV-HRG and Cross-HRG with the parameters listed in table \ref{fit-2b} for this list. It is clear how the combined effect of extra resonances and EV parameters extracted from lattice QCD improves the description of particle yields at both energies with respect to PDG2014. Furthermore Cross-EV systematically pushes the $\chi^2$ closer to the value of 1. Besides the general improvement due to the QM list \cite{mio_forthcoming}, EV effects play a relevant role for the suppression of (anti-)proton yields, being a candidate to explain the so called proton-anomaly. A similar conclusion can be extended to $\Xi^{\pm}$, for which QM exhibit an overabundance when compared to dedicated observables from lattice QCD \cite{Alba:2017mqu}, being compatible with the anomalies seen in particle yields. Charm degrees of freedom have not been included in the present study in order to be consistent with lattice observables, but they could be relevant in the feed-down of lighter particles as pions and kaons \cite{Andronic:2016nof}.\\
Nuclei are not included in the fits due to their scarce relevance in thermal fits, their production being almost independent from the particle list; i.e. the fit of $d$($\bar d$), $He^3$($\bar{He^3}$) \cite{Adam:2015vda} and $H^3_{\Lambda}$($\bar{H^3_{\Lambda}}$) \cite{Adam:2015yta} gives a temperature of $160.7\pm7.2$ MeV.\\
In tables \ref{fo-alice} and \ref{fo-star} the extracted freeze-out parameters are summarised. As already pointed out in \cite{Chatterjee:2017yhp,mio_forthcoming} the inclusion of extra higher-mass resonances decreases the freeze-out temperature; this remains true with EV effects, while on the other hand the system volume increases due to the finite sizes of particles, and partially to the larger number of states involved. Baryon chemical potential shows to be rather stable with respect to both effects.\\
It should be noted that with parameters extracted from lattice QCD there is not the second minimum structure which was found in \cite{Vovchenko:2015cbk}, stabilising the freeze-out temperature at values which are reasonably in agreement with the pseudo-critical temperature found on lattice.\\
As a final remark it is worth to say that when fitting ALICE particle yields with PDG lists and EV effects, it is possible to achieve an almost perfect description in the \emph{s-inv} scheme \cite{Alba:2016hwx}. Due to the consistency of this result with the findings presented in this paper, the success of the \emph{s-inv} against the \emph{2b} for PDG lists could be attributed to the missing strange baryons.

\begin{table}[h]
\begin{tabular}{|c|c|c|c|c|c|}
\hline
 & $\chi^2$& T (MeV)& $\mu_B$ (MeV)& V (fm$^3$)\\
\hline
PDG14 - id &31.06/10& 155.7 $\pm$ 2.2 & 1.4 $\pm$ 6.0 & 4240.1 $\pm$ 555.7 \\
\hline
QM - ExVol &18.55/10& 150.5 $\pm$ 1.7 & 5.2 $\pm$ 6.8 & 6589.4 $\pm$ 601.6 \\
\hline
QM - Cross &17.34/10& 150.9 $\pm$ 1.7 & 5.0 $\pm$ 6.7 & 6285.9 $\pm$ 600.7 \\
\hline
\end{tabular}
\caption{Results from fits to ALICE data for PDG2014 in the ideal HRG and QM in the EV-HRG and Cross-HRG with the corresponding parameters listed in table \ref{fit-2b}.}
\label{fo-alice}
\end{table}

\begin{table}[h]
\begin{tabular}{|c|c|c|c|c|c|}
\hline
 & $\chi^2$& T (MeV)& $\mu_B$ (MeV)& V (fm$^3$)\\
\hline
PDG14 - id &14.06/8& 162.3 $\pm$ 2.3 & 29.0 $\pm$ 8.2 & 1666.3 $\pm$ 211.9 \\
\hline
QM - ExVol &13.58/8& 155.6 $\pm$ 1.7 & 30.7 $\pm$ 8.6 & 2878.2 $\pm$ 243.5 \\
\hline
QM - Cross &12.08/8& 156.2 $\pm$ 1.7 & 30.8 $\pm$ 8.5 & 2726.5 $\pm$ 239.9 \\
\hline
\end{tabular}
\caption{Results from fits to STAR data for PDG2014 in the ideal HRG and QM in the EV-HRG and Cross-HRG with the corresponding parameters listed in table \ref{fit-2b}.}\label{fo-star}
\end{table}

\section{Conclusions}

In the present paper we studied the balance between attractive and repulsive interactions in lattice QCD thermodynamics by employing unmeasured higher-mass resonances and EV effects. We showed how PDG lists are systematically incomplete in the strange baryon sector, with the need for the inclusion of repulsive interactions.\\
Other than the presence of EV effects, one of the main consequences resulting from the fit to lattice data is the systematically smaller effective sizes of strange hadrons with respect to light ones with equal masses. The best description is achieved through the combined effect of QM states and a mass dependent eigenvolume with a different proportionality between light and strange sectors, thus being compatible with the available experimental measurements of the charge radii of ground state hadrons. This result could further be tested against hadrons with multiple strange quarks and with charm degrees of freedom, which in principle should be more localised and for which lattice data are already available.\\
Here we show also how the extracted parameterisation systematically improves the description of particles yields measured by ALICE and STAR experiments in the region of small $\mu_B$, enforcing the link between theory and experiment. Therefore EV effects naturally emerge as a candidate to explain the anomalies found for the proton and other particles.\\
Furthermore we show how the non-monotonic behaviour of observables on the lattice is a direct consequence of repulsive interactions without any manifest criticality, for which one would need specific attractive terms as in the vdW-HRG. We think that these attractive terms are mostly relevant for the nuclear matter region of the phase diagram, while the correct behaviour of two-particles attractive channels at $\mu_B\simeq0$ is properly accounted by the inclusion of resonances. In effect such attractive terms would hardly survive at temperatures typical of lattice simulations, since they can be connected to the presence of states with baryon number equal or larger than 2 with very small binding energies.\\
We also pointed out how the extracted parameterisation naturally involves finite sizes for mesons too, which in the vdW-HRG are treated as point-like non-interacting objects. This difference could be relevant for observables connected to net-electric charge as the $\chi_4^Q/\chi_2^Q$, which in our calculations shows the typical suppression of similar quantities calculated on the lattice, being this useful also for future ALICE measurements.\\
All the information extracted by means of EV-HRG can be used as indications for the S-matrix approach \cite{Huovinen:2017ogf,Lo:2017lym}, in order to compensate the missing information on measured phase shifts especially in the strange sector.\\
The procedure here employed is totally general, and can be repeated with a new set of lattice observables in order to better extract information on new physics, considering also the daily improvements in the precision of lattice simulations. In particular combinations of conserved charges which could be more sensitive to differences between EV-HRG and Cross-HRG can be found.\\
Temperature dependent effective masses for hadrons have shown to have interesting implications on fluctuations of conserved charges measured on the lattice \cite{Aarts:2017rrl,Aarts:2017iai}. Since they are nothing but another way to account for effective interaction, it would be interesting to clarify their overlap with the other phenomena here presented, as well as to investigate the possible mutual implications in order to better understand the nature of the physics encoded in lattice calculations.\\
The EV-HRG can be easily used to study moments of multiplicity distributions measured by STAR in order to clearly extract signals for the true CEP connected with deconfinement transition, without the contamination of other criticalities as the one deriving from the liquid-phase transition.

\vspace{0.1cm}
\noindent
{\bf Acknowledgements.}
Fruitful discussions with H. Stoecker, V. Vovchenko and C. Hegner are gratefully acknowledged. We thank F. Karsch for providing lattice data for $\chi_{31}^{BS}/\chi_{11}^{BS}$, $\chi_{31}^{BQ}/\chi_{11}^{BQ}$ and $\chi_{11}^{BQ}/\chi_2^B$. PA is grateful for support from COST Action CA15213 THOR.

\bibliography{cites}{}

\end{document}